\documentclass[runningheads]{llncs}
\usepackage{graphicx}
\usepackage{tcolorbox}
\usepackage{multirow}
\usepackage{boldline}
\usepackage[ruled,linesnumbered]{algorithm2e}
\usepackage{amssymb,amsmath}
\usepackage{hyperref}
\hypersetup{
    colorlinks,
    citecolor=blue,
    filecolor=black,
    linkcolor=red,
    urlcolor=green
}
\begin{document}
\title{Profit Maximization using Social Networks in Two\mbox{-}Phase Setting  \thanks{The work of Dr. Suman Banerjee is supported by the Start Up Grant provided by the Indian Institute of Technology Jammu, India (Grant No.: SG100047).}}
%
%
\author{Poonam Sharma \inst{1} \and
Suman Banerjee \inst{1} }
\authorrunning{Sharma and Banerjee}
%
\institute{Department of Computer Science and Engineering, \\
Indian Institute of Technology Jammu,\\
Jammu \& Kashmir-181221, India. \\ 
\email{2021rcs1023@iitjammu.ac.in, suman.banerjee@iitjammu.ac.in }
}
\maketitle              
\begin{abstract}
Now-a-days, \emph{Online Social Networks} have been predominantly used by  commercial houses for viral marketing where the goal is to maximize profit. In this paper, we study the problem of Profit Maximization in the two\mbox{-}phase setting. The input to the problem is a \emph{social network} where the users are associated with a cost and benefit value, and a fixed amount of budget splitted into two parts. Here, the cost and the benefit associated with a node signify its incentive demand and the amount of benefit that can be earned by influencing that user, respectively. The goal of this problem is to find out the optimal seed sets for both phases such that the aggregated profit at the end of the diffusion process is maximized. First, we develop a mathematical model based on the \emph{Independent Cascade Model} of diffusion  that captures the aggregated profit in an \emph{expected} sense. Subsequently, we show that selecting an optimal seed set for the first phase even considering the optimal seed set for the second phase can be selected efficiently, is an $\textsf{NP}$-Hard Problem. Next, we propose two solution methodologies, namely the \emph{single greedy} and the \emph{double greedy} approach for our problem that works based on marginal gain computation. A detailed analysis of both methodologies has been done to understand their time and space requirements. We perform an extensive set of experiments to demonstrate the effectiveness and efficiency of the proposed approaches with real-world datasets. From the experiments, we observe that the proposed solution approaches lead to more profit compared to the baseline methods and in particular, the double greedy approach leads to up to $5 \%$ improvement compared to its single\mbox{-}phase counterpart.
\end{abstract}
\section{Introduction}
In recent times, Online Social Networks has emerged as an additional dimension of human life. Among many one of the important phenomena of online social networks is the \emph{diffusion of information} using which information propagates from one part of the network to the other \cite{guille2013information}. People's decisions from personal to political are influenced at least to some extent by friends and neighbors. How we can make use of this influence in the context of viral marketing such that the profit can be maximized remains an active area of research \cite{gao2022adaptive,lu2012profit}.   

\paragraph{\textbf{Problem Background}} Commercial houses use online social networks for viral marketing purposes \cite{domingos2005mining}. The goal here is to identify a small set of influential users to activate initially that leads to maximum influence (in turn profit). These initially active nodes are called \emph{seed nodes}. In particular, given a social network and positive integer $k$ the problem is to choose a subset of $k$ nodes to maximize the influence in the network \cite{kempe2003maximizing}. This problem remains an active area of research in  the domain of social network analysis \cite{banerjee2020survey}. However, it is important to observe that from the commercial house perspective, the key objective is to maximize the profit rather than influence. 

\paragraph{\textbf{Motivation}} The traditional seed set selection methodologies for the influence and profit maximization problem consider that all the seed nodes will be deployed at one go before the diffusion process starts. However, some recent studies show that instead of one go if we split the budget into two or more parts and conduct the diffusion process in two or more rounds, then the number of influenced nodes increases \cite{dhamal2016information,sun2018multi}. The two\mbox{-}phase setting in case of influence maximization works in the following way. Suppose a commercial house wants to select $k$ many seed nodes. The first task is to divide the budget into two parts $k_1$ and $k_2$ (such that $k_1 + k_2 \leq k$) where $k_1$ (similarly $k_2$) many seed nodes will be chosen for the first (similarly second) phase. Now, $k_1$ many seed nodes are chosen to start the first phase and observe the diffusion process up to a pre\mbox{-}specified number of rounds (assume $d$\mbox{-}th round). At the end of $d$\mbox{-}th round, we deploy $k_2$ many seed nodes, and all the recently activated nodes at the $d$-th time step along with $k_2$ many nodes works as the seed set for the second phase. When the diffusion process ends after the second phase, we compute the number of influenced nodes. The same idea can be extended for the multi\mbox{-}phase case as well. It has been shown in the literature if we adopt this kind of strategy, the number of influenced nodes at the end increases over the single\mbox{-}phase diffusion approach. Naturally, a question arises, does the same thing happen even for the Profit Maximization Problem. In this paper, we elaborately investigate this question.
\paragraph{\textbf{Related Work}} The problem of profit/ benefit/ revenue maximization in the context of viral marketing has been studied extensively\cite{lu2012profit,tang2018towards,banerjee2022budgeted,banerjee2021earned}. Lu and Lakshmanan \cite{lu2012profit} was the first to study and introduce the profit maximization problem. They extended the classical linear threshold model considering the price and valuation of the product during decision making. They proposed three different algorithms for solving this problem. Tang et al. \cite{tang2018towards} proposed a two\mbox{-}phase framework for this problem and showed that it is a challenging problem as the function is neither monotone nor submodular. Gao et al. \cite{gao2022adaptive} studied the profit maximization problem in adaptive setting where the seed nodes are chosen one by one. This problem leads to the profit function which is non-adaptive submodular. Subsequently,
they developed a solution methodology that provides a data-dependent approximation guarantee. Du et al. \cite{du2021nonsubmodular} studied the same problem and showed that the function is non-submodular and it is a difference between two submodular functions. Chen et al. studied this problem in multiple product setting and called it as Profit Maximization with Multiple Adoptions Problem. They proposed an approach called reverse influence sampling which achieves $(1-\frac{1}{e}-\epsilon)$-factor approximation guarantee. Also, there are several other studies on this problem \cite{gao2019robust,salavati2019identifying}. 
\paragraph{\textbf{Our Contributions}} We study the problem of profit maximization in the two\mbox{-}phase setting. The input to the problem is a social network where the users are associated with cost and benefit value and the budgets for both the phases. Here the cost signifies the amount of incentive need to be paid to that user to include him/her as a seed node. The benefit value signifies that the benefit that can be earned from that user if (s)he is influenced. The goal is to choose  optimal seed sets for both the phases such that the aggregated profit is maximized. In particular, we make the following contributions in this paper:
\begin{itemize}
\item We study the \textsc{Profit Maximization Problem} using social networks in the two\mbox{-}phase setting. To the best of our knowledge, this is the first study on profit maximization in this direction.
\item We develop a mathematical model for this problem that captures the expected profit at the end of the diffusion process. Subsequently, we show that selecting an optimal seed set for the first phase even considering the optimal seed set for the second phase can be selected efficiently is an $\textsf{NP}$-Hard Problem.
\item We propose two algorithms, namely the single greedy and the double greedy approach along with their detailed analysis and both of them work based on marginal profit gain computation. 
\item Finally, we conduct an extensive set of experiments with real\mbox{-}world datasets to show that the proposed methodologies can lead to more amount of profit than the baseline methods.
\end{itemize}
\paragraph{\textbf{Organization of the Paper}} The rest of the paper is organized as follows. In Section \ref{Sec:Prob_Def}, we describe the background and define the problem formally. Section \ref{Sec:Prop_Method} contains the proposed solution methodologies. Section \ref{Sec:EE} contains the experimental evaluation of the solution approaches. Finally, Section \ref{Sec:CFD} concludes our study. 
\section{Background and Problem Definition} \label{Sec:Prob_Def}
We represent the input social network by a simple (un)directed, and edge\mbox{-}weighted graph denoted by $G(V, E, \mathcal{P})$. Here, $V(G)=\{u_1, u_2, \ldots, u_n\}$ are the set of $n$ users and $E(G)=\{e_1, e_2, \ldots, e_m\}$ are the set of $m$ social ties. $\mathcal{P}$ denotes the edge weight function that maps each edge to its corresponding influence probability; i.e.; $\mathcal{P}:E(G) \longrightarrow (0,1]$. For any edge $(uv) \in E(G)$, let $\mathcal{P}_{uv}$ denotes the influence probability of the user $u$ on $v$. If $(uv) \notin E(G)$ then $\mathcal{P}_{uv}=0$. Each user of the network is associated with a cost and benefit value that are characterized by the cost and the benefit function denoted as $C$ and $b$. Hence, $C:V(G) \longrightarrow \mathbb{Z}^{+}$ and $b:V(G) \longrightarrow \mathbb{Z}^{+}$.  For any $u \in V(G)$, let $C(u)$ and $b(u)$ denote the cost and benefit associated with the user $u$.
\par To conduct the diffusion process in the network, a subset of the users are chosen as a seed user and they are considered to be influenced at time step $t=0$. Now, the information is diffused in the network based on some rules. In this paper, we consider that the information in the network is diffused by the rule of the Independent Cascade Model. By this model, (i) information is diffused in discrete time steps, (ii) a node can be either of the following two states: `uninfluenced' and `influenced', (iii) every influenced node at time step $t$ will get a single chance to activate its inactive neighbor, and finally, (iv) a node can change its state from `uninfluenced' to  `influenced', however not the vice versa, (v) diffusion process ends when no more node activation is possible. The diffusion process can be expressed as a \emph{live graph} and they are $2^{m}$ many where $m$ denotes the number of edges in $G$. We denote these graphs as $L(G)=\{ \mathcal{G}_{1}, \mathcal{G}_{2}, \ldots, \mathcal{G}_{2^{m}}\}$. Due to the space limitation, we are unable to discuss more on live graphs, however, it can be found in \cite{kempe2003maximizing}. For the input social network $G$, we denote its live graphs by $L(G)$. Next, we state the influence of a seed set in Definition \ref{Def:1}. 
\begin{definition}[Influence of a Seed Set] \label{Def:1}
Assume that $\mathcal{S} \subseteq V(G)$ is the seed set. Now, at the end of the diffusion process starting from $\mathcal{S}$, the number of nodes that are influenced is called the influence of $\mathcal{S}$. We denote this by $\sigma(\mathcal{S})$ where $\sigma()$ is the social influence function that maps each subset of the user to their corresponding influence value, i.e., $\sigma: 2^{V(G)} \longrightarrow \mathbb{R}_{0}$ with the condition $\sigma(\emptyset)=0$. 
\end{definition}
Now, one natural question arises in this context is which nodes should be chosen as seed nodes to maximize the influence. Formally, this problem is called as the Social Influence Maximization Problem stated in Definition \ref{Def:2}.

\begin{definition}[Social Influence Maximization Problem] \label{Def:2}
Given a social network $G(V, E, \mathcal{P})$, and a positive integer $k$ the goal of the social influence maximization problem is to choose a subset of $k$ nodes $\mathcal{S} \subseteq V(G)$ such that their initial activation leads to maximum number of influenced nodes. Mathematically, this problem can be stated as follows:
\begin{equation}
\mathcal{S}^{OPT} = \underset{\mathcal{S} \subseteq V(G) \text{ and }|\mathcal{S}| \leq k}{argmax} \ \sigma(\mathcal{S})
\end{equation} 
\end{definition}
Here, $\mathcal{S}^{OPT}$ denotes an optimal seed set of size $k$. Now, as mentioned previously, in case of viral marketing of commercial products the main concern is profit rather than influence. Profit of a seed set is defined as the benefit earned by the seed set minus the total cost of the seed set. Hence, to define the profit by a seed set, we should first define the benefit earned by a seed set which is stated in Definition \ref{Def:3}. 
  \begin{definition}[Benefit Earned by a Seed Set] \label{Def:3}
  Given a social network $G(V, E, \mathcal{P})$ and a seed set $\mathcal{S}$ we denote the benefit obtained by $\mathcal{S}$ as $\beta(\mathcal{S})$ and defined in terms of expectation over the set of all possible live graphs. Mathematically, this can be defined using Equation \ref{Eq:2}.
  \begin{equation} \label{Eq:2}
  \beta(\mathcal{S})= \mathbb{E} \ [\underset{v \in I_{\mathcal{G}_{i}}(\mathcal{S}) \ \cap \ V(G)}{\sum} \ b(v) \ ] 
  \end{equation}
  Here, $I_{\mathcal{G}_{i}}(\mathcal{S})$ denotes the set of influenced nodes from the seed set $\mathcal{S}$ in the $i$-th live graph. $C(\mathcal{S})$ denotes the total cost of the seed set, i.e., $C(\mathcal{S})= \underset{v \in \mathcal{S}}{\sum} C(v)$.
  \end{definition}
  The expectation mentioned in Definition \ref{Eq:2} is taken over the probability distribution of the benefit values in different live graphs. Now, once we have the benefit value for a seed set and we can calculate the cost of a seed set, we can immediately define the profit of the seed set by subtracting its cost from its benefit. For a seed set $\mathcal{S}$, we denote its profit by $\phi(\mathcal{S})$ and can be defined using Equation \ref{Eq:Profit}.

\begin{equation} \label{Eq:Profit}
\phi(\mathcal{S})= \beta(\mathcal{S}) \ - \ C(\mathcal{S}) 
\end{equation}  
 Based on our definition of profit of a seed set in Equation \ref{Eq:Profit}, now state the profit maximization problem in Definition \ref{Def:PMP}.
 
 \begin{definition}[Profit Maximization Problem] \label{Def:PMP}
 Given a social network \\ $G(V,E, \mathcal{P})$ where users of the network are associated with cost and benefit value and a fixed amount of budget $\mathcal{B}$ is given. The goal is to choose a subset of the nodes $\mathcal{S} \subseteq V(G)$ such that the profit is maximized. Mathematically, this problem can be stated as follows:
 \begin{equation}
 \mathcal{S}^{OPT}=\underset{\mathcal{S} \subseteq V(G) \text{ and }C(\mathcal{S}) \leq \mathcal{B}}{argmax} \ \phi(\mathcal{S})
 \end{equation}
 \end{definition}   
  As mentioned previously, instead of deploying all the seeds in one go we can split the budget in two or more parts and conduct the diffusion process accordingly. In this paper, we consider the two\mbox{-}phase case and define the \textsc{Two\mbox{-}Phase Profit Maximization} Problem which is stated in Definition \ref{Def:TPPMP}.

\begin{definition}[Two\mbox{-}Phase Profit Maximization Problem] \label{Def:TPPMP}
Given a social network $G(V,E, \mathcal{P})$ where uses of the network are associated with cost and benefit value and budget for two phases $\mathcal{B}_{1}$ and $\mathcal{B}_{2}$ our goal is to choose seed nodes $\mathcal{S}_{1}$ and $\mathcal{S}_{2}$ such that $C(\mathcal{S}_{1}) \leq \mathcal{B}_{1}$ and $C(\mathcal{S}_{2}) \leq \mathcal{B}_{2}$ and the goal is to maximize the profit at the end of second phase. 
\end{definition} 
From the computational point of view this problem can be posed as follows:

\begin{center}
\begin{tcolorbox}[title=\textsc{Influential Billboard Slot Selection Problem}, width=12.5cm]
\textbf{Input:} A social Network $\mathcal{G}(\mathcal{V},\mathcal{E}, \mathcal{P})$, with cost and benefit function $\mathcal{C}$ and $b$, respectively and the budget $\mathcal{B}$ with their split $\mathcal{B}_1$ and $\mathcal{B}_2$.

\textbf{Problem:} Find out optimal seed sets for first and second phase $\mathcal{S}_{1}$ and $\mathcal{S}_{2}$, respectively for the diffusion. 
\end{tcolorbox}
\end{center}  
   
 In this paper, we formulate a mathematical model and two solution methodologies for this problem which are presented in the subsequent section. 
   
\section{Mathematical Model and Solution Methodologies} \label{Sec:Prop_Method}
In this section, we describe our mathematical model and the proposed solution methodologies. Initially, we start by describing  
\paragraph{\textbf{Mathematical Model}} Let, a live graph $\mathcal{G} \in L(G)$ with its generation probability $P(\mathcal{G})$ is destined to occur and $\mathcal{S}_{1}$ be the seed set for the Phase I. The diffusion process starts on a live graph $\mathcal{G}$ with seed set $\mathcal{S}_{1}$ by the rule of IC Model and observe this diffusion process till time step $d$. Then, we will have the information regarding which nodes are influenced and which are not. We call this as the partial observation till time step $d$ and denoted as $Y$. So, at the end of time step $d$, we have already activated nodes denoted by $A_{Y}$ and newly activated nodes (at time step $d$) $R_{Y}$. These two sets $A_{Y}$ and $R_{Y}$ are determined from the partial observation $Y$. 
\par Now, as we have the partial observation $Y$, for a subset of the edges of the live graph $\mathcal{G}$ we are sure whether they have appeared or not and based on that we can update the generation probability $P(\frac{\mathcal{G}}{Y})$. Now, the second phase needs to begin, and assume that $\mathcal{S}^{OPT(Y,\mathcal{B}_2)}_{2}$ denotes the optimal seed set for Phase II when the partial observation $Y$ and the budget is $\mathcal{B}_2$. At the time step $d$, we deploy the nodes in the set $\mathcal{S}^{OPT(Y,\mathcal{B}_2)}_{2}$ along with the nodes in $R_{Y}$  both of them together will act as seed set for Phase II. Now, our goal is to calculate the expected profit that can be earned in Phase II. Now, it is important to
observe that in Phase II , the nodes from which the profit can be earned will be the subset from $V(G) \setminus A_{Y}$. So, for the given partial observation $Y$ (hence, newly activated nodes $R_Y$) along with an optimal seed set for Phase II, i.e., $\mathcal{S}^{OPT(Y,\mathcal{B}_2)}_{2}$ will be equals to $\displaystyle{\sum_{\mathcal{G} \in L(G)}P(\frac{\mathcal{G}}{Y}})[\phi^{V(\mathcal{G})\setminus A_{Y}}(R_{Y} \cup S_{2}^{OPT(Y,B_{2})})]$. Here, $\phi^{V(\mathcal{G})\setminus A_{Y}}(\mathcal{S})$ denotes the profit earned by the seed set $\mathcal{S}$ from the graph $V(G) \setminus A_{Y}$.
\par We can observe that given a live graph $\mathcal{G}$, seed set of Phase I; i.e.; $\mathcal{S}_{1}$, and the time step $d$, we can get the partial observation $Y$. Hence, $\mathcal{S}^{OPT(Y,\mathcal{B}_2)}_{2}$ can be written as $S_{2}^{OPT(X, S_{1},d,\mathcal{B}_{2})}$. Now, to develop an objective function where the decision variable will be the seed set for Phase I, we assume that given the partial observation $Y$, we will select an optimal seed set for Phase II. It is important to observe that at the starting of Phase I, the partial observation $Y$ is not known. Let our objective function be $\mathbb{F}(S_1, d, \mathcal{B}_2 )$ as the expected profit with respect to all possible occurrences $Y$. Assuming that $d$ and $\mathcal{B}_2$ are already given, so we can write $\mathbb{F}(\mathcal{S}_{1} , d, \mathcal{B}_{2} )$ as $f(\mathcal{S}_{1})$. In the following derivation, by $\mathcal{S}^{'}_{2}$ we denote the set $R_{Y} \cup S_{2}^{OPT(Y,B_{2})}$. 
\begin{center}
		$f(S_{1}) = \displaystyle{\sum_{Y}P({Y}})\Bigg\{{\Big\{ \displaystyle{\sum_{\mathcal{G}}P(\frac{\mathcal{G}}{Y}})\phi(A_{Y}) + \displaystyle{\sum_{\mathcal{G}}P(\frac{\mathcal{G}}{Y}})[\phi^{V(\mathcal{G})\setminus A_{Y}}(\mathcal{S}^{'}_{2})]}\Big\}\Bigg\}$\\
		$= \displaystyle{\sum_{Y}P({Y}})\displaystyle{\sum_{\mathcal{G}}P(\frac{\mathcal{G}}{Y}})\Bigg\{{\Big\{\phi(A_{Y}) + [\phi^{V(\mathcal{G})\setminus A_{Y}}( S_{2}^{OPT(X,S_{1},d,\mathcal{B}_{2})})]}\Big\}\Bigg\}$\\
		$= \displaystyle{\sum_{Y}P({Y}})\displaystyle{\sum_{\mathcal{G}}P(\frac{\mathcal{G}}{Y}})\Bigg\{{\phi^{\mathcal{G}}(S_{1} \cup S_{2}^{OPT(X,S_{1},d,\mathcal{B}_{2})})}\Bigg\}$\\
		$\Bigg\{ \because \displaystyle{\sum_{Y}P({Y}})\displaystyle{\sum_{\mathcal{G}}P(\frac{\mathcal{G}}{Y}}) = \displaystyle{\sum_{Y}}\displaystyle{\sum_{\mathcal{G}}} {\frac{P(\mathcal{G},Y)}{P(Y)}} P(Y)$\\
		$ = \displaystyle{\sum_{Y}}\displaystyle{\sum_{\mathcal{G}}} P(\mathcal{G},{Y})$ \ 
		$ = \displaystyle{\sum_{\mathcal{G}}}\displaystyle{\sum_{Y}} P(\mathcal{G},{Y})$\
		$= \displaystyle{\sum_{\mathcal{G}}} P(\mathcal{G})\Bigg\}$\\
\end{center}
The objective function formulated of our Two\mbox{-}Phase Profit Maximization Problem is as follows:
	\begin{equation}
		\label{Eq: Obj_eq}
		\therefore f(\mathcal{S}_{1})= \displaystyle{\sum_{\mathcal{G}}}P(\mathcal{G}){\phi^{\mathcal{G}}(\mathcal{S}_{1} \cup \mathcal{S}_{2}^{OPT(X,\mathcal{S}_{1},d,\mathcal{B}_{2})})}
	\end{equation}
	Now, it is important to observe that the developed model considers the optimal seed set selection in Phase II, which is itself an \textsf{NP}-Hard problem. So, Theorem \ref{Th:1} holds.
	
	\begin{theorem} \label{Th:1}
	Finding the optimal seed set $\mathcal{S}_{1}$ that maximizes $f(\mathcal{S}_{1})$ as mentioned in Equation \ref{Eq: Obj_eq} is \textsf{NP}-Hard.
\end{theorem}
Now, we list out few important properties of the objective function. Due to the space constraint, we are not able to provide the proofs.

\begin{lemma}
$f (.)$ may be positive or negative.
\end{lemma}

\begin{lemma}
$f (.)$ is neither monotonically increasing nor monotonically decreasing.
\end{lemma} 
	 
	\begin{lemma}
$f (.)$ is neither submodular nor supermodular.
\end{lemma}  
\begin{lemma} \label{Lemma:4}	 
$f (.)$ is neither subadditive nor superadditive.	 
\end{lemma}	 
\paragraph{\textbf{Solution Methodologies}} In this section, we describe two  solution methodologies, namely Single Greedy and Double Greedy for the Two\mbox{-}Phase Profit Maximization Problem and both of them are based on Marginal Profit Gain which is stated in Definition \ref{Def:6}.
\begin{definition} [Marginal Profit Gain] \label{Def:6}
Given a social network $G(V, E, \mathcal{P})$, a seed set $S$, and a node $u \in V(G) \setminus S$ we denote the Marginal Profit Gain for the node $u$ with respect to the seed set $S$ as $\phi_u(S)$ and it is defined as the difference of profit earned when $u$ is added to $S$ and when $u$ is not in $S$. As shown in Lemma \ref{Lemma:4}, the profit function $\phi()$ may be non-monotone as well. As we are considering the `gain', for any seed set $S$ and node $u \in V(G) \setminus S$, $\phi_u(S)$ is defined only when $\phi_u(S) > \phi(S)$. Mathematically, this can be defined using Equation \ref{Eq:Marginal_Profit_Gain}.

\begin{equation} \label{Eq:Marginal_Profit_Gain}
			\phi_u(S) = \phi(S \cup \{u\}) - \phi(S) \quad such \; that \; \phi_u(S) > \phi(S)
		\end{equation}
\end{definition}
\paragraph{\textbf{Single Greedy Approach}} Now, we describe the single greedy algorithm in two\mbox{-}phase setting. We have the following inputs: the social network $G(V, E, \mathcal{P})$, Budgets for both the phases $\mathcal{B}_{1}$ and $\mathcal{B}_{2}$, and the duration of Phase I which is $d$. Now, in the first phase, until the budget $\mathcal{B}_1$ is exhausted, we iteratively select seed node based on the marginal profit gain. In each iteration, for every non seed node $u$, we compute the marginal profit gain to its cost ratio and the node maximizes this quantity is found out. If the marginal profit gain of this node is strictly positive then it is included in the seed set of Phase I. It may so happen that
the allocated budget for the first phase has not been exhausted totally. If so, the remaining budget of Phase I is added to Phase II. Now, we conduct  diffusion process based on the IC Model starting from the seed set $\mathcal{S}_1$ till $d$-th time step. Thus, at the end of first Phase we have $A_Y$ and $R_Y$ as the influence of $\mathcal{S}_1$ at time step $d$. Now, we begin Phase II with the updated budget. The process of seed set selection is quite similar to the first phase with one difference.
During the first phase we are dealing with the entire social network, and hence, while computing the marginal profit gain we consider the whole network. However, for the second phase we deal with the network obtained by deleting the already activated nodes from the original network. Accordingly, during the second phase while computing the marginal profit gain we consider the remaining network. Algorithm \ref{Algo: Algorithm1_Simple Greedy} shows the pseudocode of the proposed approach.

\begin{algorithm}[h!]
\SetAlgoLined
\KwData{$G, B_{1}, B_{2}, d$}
\KwResult{  $S$}
 Initialize $S \leftarrow \emptyset$ \;
 \textbf{Seed set selection for First Phase}\;
 Initialize $S_{1} \leftarrow \emptyset$\;
 \While{$\texttt{TRUE}$}{
 Find $u' \leftarrow {argmax}_{u \in V(G) \setminus S_{1}} {\frac{\phi(S_{1} \cup \{u\}) - \phi(S_{1})}{C(u)}}$\;
 \If{$(\phi_{u'}(S_{1})) \leq 0$}{
 Break\;
 }
 \If{$C(u^{'}) \leq B_{1}$}{
 $S_{1} \longleftarrow S_{1} \cup \{u'\}$; $B_{1} \longleftarrow B_{1} - C(u^{'})$\;
 
 }
 }
Return $S_{1}$\;
From the partial observation in $G$ using seed set $S_{1}$ at time step $d$, we have recently activated nodes, $R_{Y}$ and already activated nodes  $A_{Y}$\;

\textbf{Seed set selection for Second Phase}\;

Initialize $S_{2} \leftarrow \emptyset$\;
Update $B_{2} \leftarrow B_{2} + B_{1}$\;
\While{$\texttt{TRUE}$}{
	
		Find $v' \leftarrow {argmax}_{v \in V(G) \setminus A_{Y}}{\frac{\phi_{V(G) \setminus A_{Y}}(S_{2} \cup \{v\}) - \phi_{V(G) \setminus A_{Y}}(S_{2})}{C(v)}}$\;
		\If{$(\phi_{v'}(S_{2})) \leq 0$}{
			Break\;
		}
		\If{$C(v^{'}) \leq B_{2}$}{
		$S_{2} \longleftarrow S_{2} \cup \{v'\}$; $B_{2} \longleftarrow B_{2} - C(v^{'})$\;
		
		}	
}
Return $S_{2}$\;
$S \longleftarrow S_{1} \cup  S_{2}$\;
Return $S$\;
 \caption{ Simple Greedy Algorithm for Two\mbox{-}Phase Profit Maximization Problem}
 \label{Algo: Algorithm1_Simple Greedy}
\end{algorithm}	

\par Next, we analyze this algorithm to understand its time and space requirement. Let, $\mathcal{C}_{min}$ denotes the minimum cost among all the nodes; i.e.; $\mathcal{C}_{min}=\underset{u \in V(G)}{min} \ C(u)$. So it is easy to observe that the maximum number of nodes that can be selected as seed in Phase I will be of $\mathcal{O}(\frac{\mathcal{B}_1}{\mathcal{C}_{min}})$ and also these many iterations are required. In each iteration, the main computation involved is the marginal profit gain. It is easy to observe that in every iteration the number of nodes for which the marginal profit gain needs to be computed is of $\mathcal{O}(n)$. Now, for a given seed node
computing the marginal gain is equivalent to traversing the graph and this takes $\mathcal{O}(m+n)$ time. In the worst case, the size of $\mathcal{S}_1$ can be of $\mathcal{O}(n)$. Hence, for one marginal gain computation time requirement is of $\mathcal{O}(n \cdot (m+n))$. As, there are $\mathcal{O}(n)$ many marginal profit gain computation, hence time requirement for this purpose is of $\mathcal{O}(n^{2} \cdot (m+n))$. Now, choosing the node that causes the maximum marginal profit gain that takes $\mathcal{O}(n)$ time. So, the time requirement for the Phase I is of $\mathcal{O}(\frac{\mathcal{B}_1}{\mathcal{C}_{min}} \cdot n^{2} \cdot (m+n))$. Now, before starting Phase II, we need to delete the already activated nodes in Phase I from the Graph. Now, the number of already activated nodes are of $\mathcal{O}(n)$. In the worst case, they may be incident with $\mathcal{O}(n^{2})$ many edges. So, deleting these vertices from the graph requires $\mathcal{O}(n^{2})$ time. The analysis of the second phase will remain the same except one difference. As in Phase II, the graph has been reduced by deleting the already activated nodes so $\mathcal{C}_{min}$ may not be minimum cost of the nodes in Phase II. Let, it be $\mathcal{C}^{'}_{min}$ and this means $\mathcal{C}^{'}_{min}=\underset{u \in V(G) \setminus A_{Y}}{min} \ C(u)$. However, the remaining computations remains the same. So, the time requirement for Phase II will be of $\mathcal{O}(\frac{\mathcal{B}_2}{\mathcal{C}^{'}_{min}} \cdot n^{2} \cdot (m+n))$. Now, summing everything up, the total time requirement of the single greedy approach will be of $\mathcal{O}((\frac{\mathcal{B}_1}{\mathcal{C}_{min}}+\frac{\mathcal{B}_2}{\mathcal{C}^{'}_{min}}) \cdot n^{2} \cdot (m+n))$. Now, the extra space consumed by this method is to store the seed sets for both the phases which can be of $\mathcal{O}(n)$. Hence, Theorem \ref{Th:1} holds.

\begin{theorem} \label{Th:1}
The time and space requirement of Single Greedy Approach is of $\mathcal{O}((\frac{\mathcal{B}_1}{\mathcal{C}_{min}}+\frac{\mathcal{B}_2}{\mathcal{C}^{'}_{min}}) \cdot n^{2} \cdot (m+n))$ and $\mathcal{O}(n)$, respectively.
\end{theorem}
It is easy to convince that two\mbox{-}phase setting is a generalization of single\mbox{-}phase setting and it has been mentioned in \cite{tang2017profit} that the single greedy approach in one\mbox{-}phase setting does not lead to any constant factor approximation guarantee. Next, we describe the double greedy approach. 
 \paragraph{\textbf{Double Greedy Approach}} First phase of this method goes like this. We initialize two sets $\mathcal{S}_{1}$ and $\mathcal{T}_{1}$. The first one is with $\emptyset$ and the second one is with $V(G)$. Now, for every node $u \in V(G)$, we compute two measures $r^{+}_{u}$ and $r^{-}_{u}$ which are mentioned in Equation \ref{Eq:7} and \ref{Eq:8}, respectively.
 
\begin{minipage}[t]{0.45\textwidth}
\begin{equation} \label{Eq:7}
r^{+}_{u} \leftarrow \frac{\phi(S_{1} \cup \{u\}) - \phi(S_{1})}{C(u)}
\end{equation}
\end{minipage}
\begin{minipage}[t]{0.45\textwidth}
\begin{equation} \label{Eq:8}
r^{-}_{u} \leftarrow \frac{\phi(\mathcal{T}_{1} \setminus \{u\}) - \phi(\mathcal{T}_{1})}{C(u)}
\end{equation}
\end{minipage} 

 Now, if $r^{+}_{u} \geq r^{-}_{u}$ and the cost of the current nodes is less than the available budget, then the set $\mathcal{S}_{1}$ is updated as $\mathcal{S}_{1} \cup \{u\}$ and the budget $\mathcal{B}_{1}$ is updated as $\mathcal{B}_{1}- \mathcal{C}(u)$, though $\mathcal{T}_{1}$ remains the same. If the budget is not sufficient or if $r^{+}_{u} < r^{-}_{u}$, $\mathcal{T}_{1}$ is reduced by deleting the current node and in that case $\mathcal{S}_{1}$ remains the same. After repeating these steps we obtain the seed set for Phase I; i.e.; $\mathcal{S}_{1}$. Now, we conduct the diffusion process and observe upto the time step $d$ and this obtain the the already activated nodes and recently activated nodes. If there is any unutilized budget of Phase I, that has been added to the budget of Phase II. For the seed set selection of Phase II, we repeat the same process however on the reduced graph; i.e.; the graph obtained by deleting the already activated nodes in the first phase. Now, we proceed to describe the analysis of the double greedy approach.
 
\begin{algorithm}[!htbp]
\SetAlgoLined
\KwData{$G, B_{1}, B_{2}, d$}
\KwResult{$S$}
 Initialize $S \leftarrow \emptyset$ \;
 \textbf{Seed set selection for First Phase}\;
 Initialize $S_{1} \leftarrow \emptyset$, $T_{1} \leftarrow V(G)$\;
 
 \For{$\text{All } u \in V(G)$}{
$ r^{+}_{u} \leftarrow \frac{\phi(S_{1} \cup \{u\}) - \phi(S_{1})}{C(u)}$\;
$ r^{-}_{u} \leftarrow \: -{\frac{\phi(T_{1} \setminus \{u\}) - \phi(T_{1})}{C(u)}}$\;
\eIf{$r^{+}_{u} \geq r^{-}_{u}$}{
\eIf{$C(u) \leq B_{1}$}{
$S_{1} \leftarrow S_{1} \cup \{u\}$; $T_{1} \: remains \: same$\;
$B_{1} \longleftarrow B_{1} - C(u)$\;
}{

$T_{1} \leftarrow T_{1} \setminus \{u\}$; $S_{1} \: remains \: same$\;
}

}
{
$T_{1} \leftarrow T_{1} \setminus \{u\}$; $S_{1} \: remains \: same$\;

}

%

}
Return $S_{1} (=T_{1})$\;
From the partial observation in $G$ using seed set $S_{1}$ at time step $d$, we have recently activated nodes, $R_{Y}$ and already activated nodes  $A_{Y}$\;

\textbf{Seed set selection for Second Phase}\;

Initialize $S_{2} \leftarrow \emptyset$,  $T_{2} \leftarrow V(G) \setminus A_{Y}$\;
Update $B_{2} \leftarrow B_{2} + B_{1}$\;

 \For{$\text{All } v \in V(G) \setminus A_{Y}$}{
 
$ r^{+}_{v} \leftarrow \frac{\phi_{V(G) \setminus A_{Y}}(S_{1} \cup \{v\}) - \phi_{V(G) \setminus A_{Y}}(S_{1})}{C(v)}$\;
$ r^{-}_{v} \leftarrow -{\frac{\phi_{V(G) \setminus A_{Y}}(T_{1} \setminus \{v\}) - \phi_{V(G) \setminus A_{Y}}(T_{1})}{C(v)}}$\;

\eIf{$r^{+}_{v} \geq r^{-}_{v}$}{
\eIf{$C(v) \leq B_{2}$}{
$S_{2} \leftarrow S_{2} \cup \{v\}$; $T_{2} \: remains \: same$\;
$B_{2} \longleftarrow B_{2} - C(v)$\;
}{

$T_{2} \leftarrow T_{2} \setminus \{v\}$; $S_{2} \: remains \: same$\;

}

}
{
$T_{2} \leftarrow T_{2} \setminus \{v\}$; $S_{2} \: remains \: same$\;

}

}
Return $S_{2} (=T_{2})$\;
$S \longleftarrow S_{1} \cup  S_{2}$\;
Return $S$\;
 \caption{ Double Greedy Algorithm for Two\mbox{-}Phase Profit Maximization Problem}
 \label{Algo: Algorithm2_Double Greedy}
\end{algorithm}

 \par The analysis is quite similar to the single greedy approach in two\mbox{-}phase setting. As stated previously, computing one marginal profit gain computation requires $\mathcal{O}(m+n)$ time. It is important to observe that in this method we are performing two marginal profit gain computations per node. Hence, the time requirement for the seed set selection of the first phase is of $\mathcal{O}(n \cdot (m+n))$. Other than the marginal gain computations, all the remaining statements from Line $7$ to $17$ will take $\mathcal{O}(1)$ time. Now, in the worst case the size of $\mathcal{S}_{1}$, $R_{Y}$, and $A_{Y}$ can be of $\mathcal{O}(n)$. So, there can be $\mathcal{O}(n^{2})$ many edges associated with the vertices of $A_{Y}$. Hence, deleting the set $A_{Y}$ leads to the modification of the $\mathcal{O}(n^{2})$ many adjacency matrix entries of the input social network. Thus, performing the deletion step after Phase I requires $\mathcal{O}(n^{2})$ time. Like Phase I, it is easy to observe that the time requirement for seed set selection in Phase II will be of $\mathcal{O}(n \cdot (m+n))$. Hence, the total time requirement for the double greedy approach in two\mbox{-}phase setting will be of $\mathcal{O}(n \cdot (m+n) + n^{2})= \mathcal{O}(n(m+n))$. The extra space consumed by this algorithm is to store the sets $S_{1}$, $T_{1}$, $S_{2}$, $T_{2}$, and $S$. In the worst case, all of them will consume $\mathcal{O}(n)$ space. Hence, Theorem \ref{Th:2} holds.
 
 \begin{theorem} \label{Th:2}
 Running time and the space requirement of the double greedy approach in the two\mbox{-}phase setting is of $ \mathcal{O}(n(m+n))$ and $ \mathcal{O}(n)$, respectively.
 \end{theorem}

\section{Experimental Evaluation}\label{Sec:EE}
In this section, we describe the experimental evaluation of the proposed solution approaches. First, we mention the datasets that we have used.

\paragraph{\textbf{Datasets} }
We have used three datasets for our experiments namely, \emph {email-Eu-core} \cite{yin2017local} \cite{leskovec2007graph}, \emph {soc-sign-bitcoin-alpha} \cite{kumar2016edge} \cite{kumar2018rev2} and \emph {wiki-Vote} \cite{leskovec2010signed} \cite{leskovec2010predicting}. All these three datasets are publicly available and have been downloaded from \url{https://snap.stanford.edu/data/} The basic statistics of these datasets is given below:
\begin{center}
\resizebox{\columnwidth}{!}{\begin{tabular}{|c | c | c | c| c| c|} 
 \hline
 Dataset Name & Type of Graph & Number of Nodes & Number of Edges & Maximum Degree & Average Degree \\ [0.5ex] 
 \hline
 email-Eu-core & Undirected & 1,005 & 16,706 & 347 & 33.25 \\ 
 \hline
 soc-sign-bitcoin-alpha & Directed & 3,783 & 24,186 & 888 & 12.79 \\
 \hline
 wiki-Vote & Directed & 7,115 & 103,689 & 1167 & 29.15 \\ [1ex] 
 \hline
 
\end{tabular}}
\end{center}

\paragraph{\textbf{Experimental Setup} }
In our experimental setup Independent Cascade Model is used for diffusion process. The influence probability $\mathcal{P}_{uv}$ is set to 0.01. The cost setting of a $u \in V(G)$ is   $C(u) \longrightarrow [50,100]$ and benefit setting is $b(u)  \longrightarrow [800,1000]$. Hence, a cost and benefit is associated for every node $u$ in the graph. These settings are followed for all these three datasets specified above. We have evaluated datasets for Budget = $\{500, 1000, 1500, 2000, 2500\}$. Each budget is  split into a ratio of $60\%$ at time step $d = 3$. The budgets $\mathcal{B}_{1}$ and $\mathcal{B}_{2}$ are consumed in first phase and second phase, respectively. The first phase of experiment is run for 100 iterations which generates 100 sets of recently active nodes. Now, the second phase runs for 100 times for all 100 sets of recently active nodes. The maximum of the profits from the 100 recently active nodes is the profit earned from the second phase.

\begin{figure*}[!ht]
\centering
\resizebox{\columnwidth}{!}{\begin{tabular}{ccc}
\includegraphics[scale=0.2]{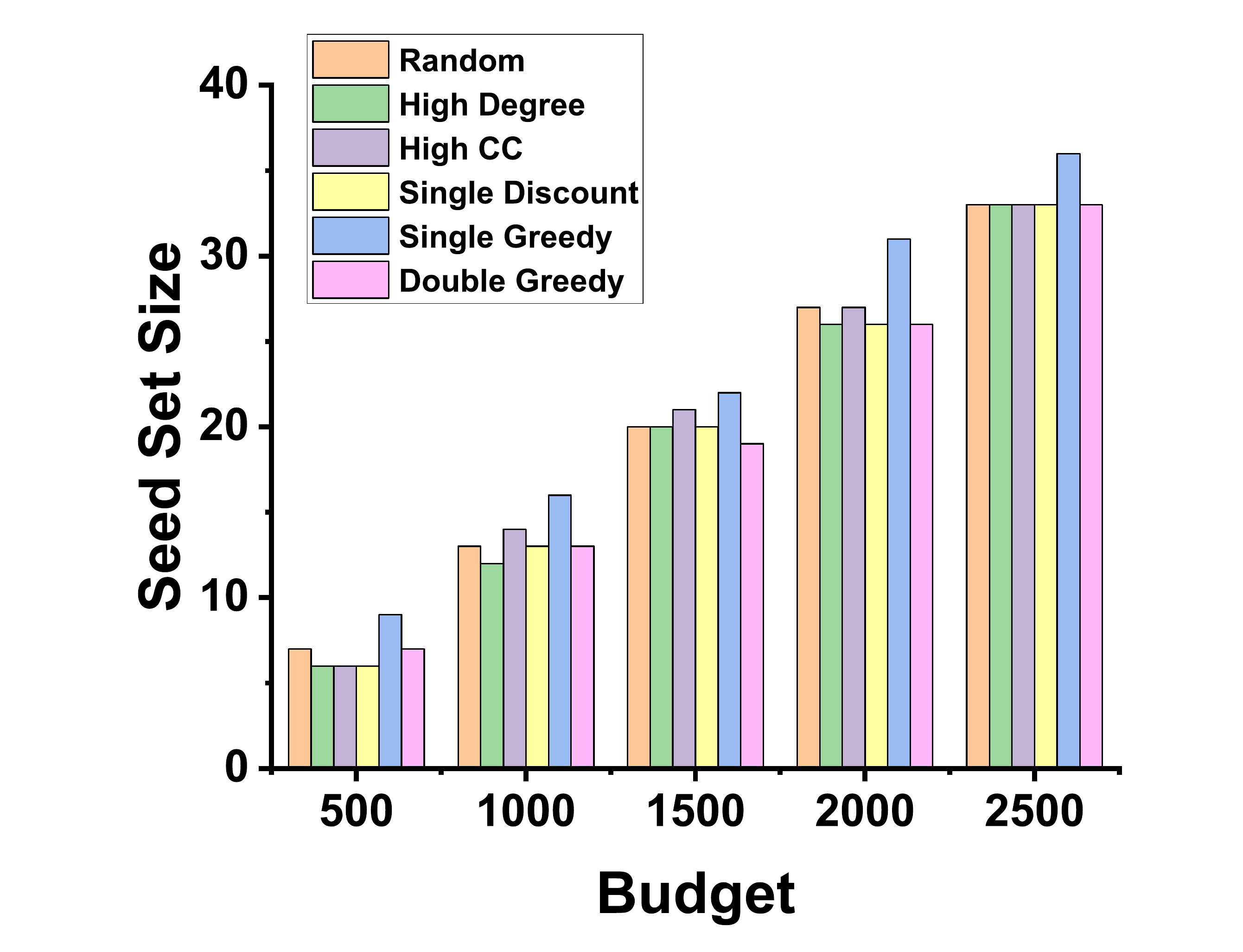} & \includegraphics[scale=0.2]{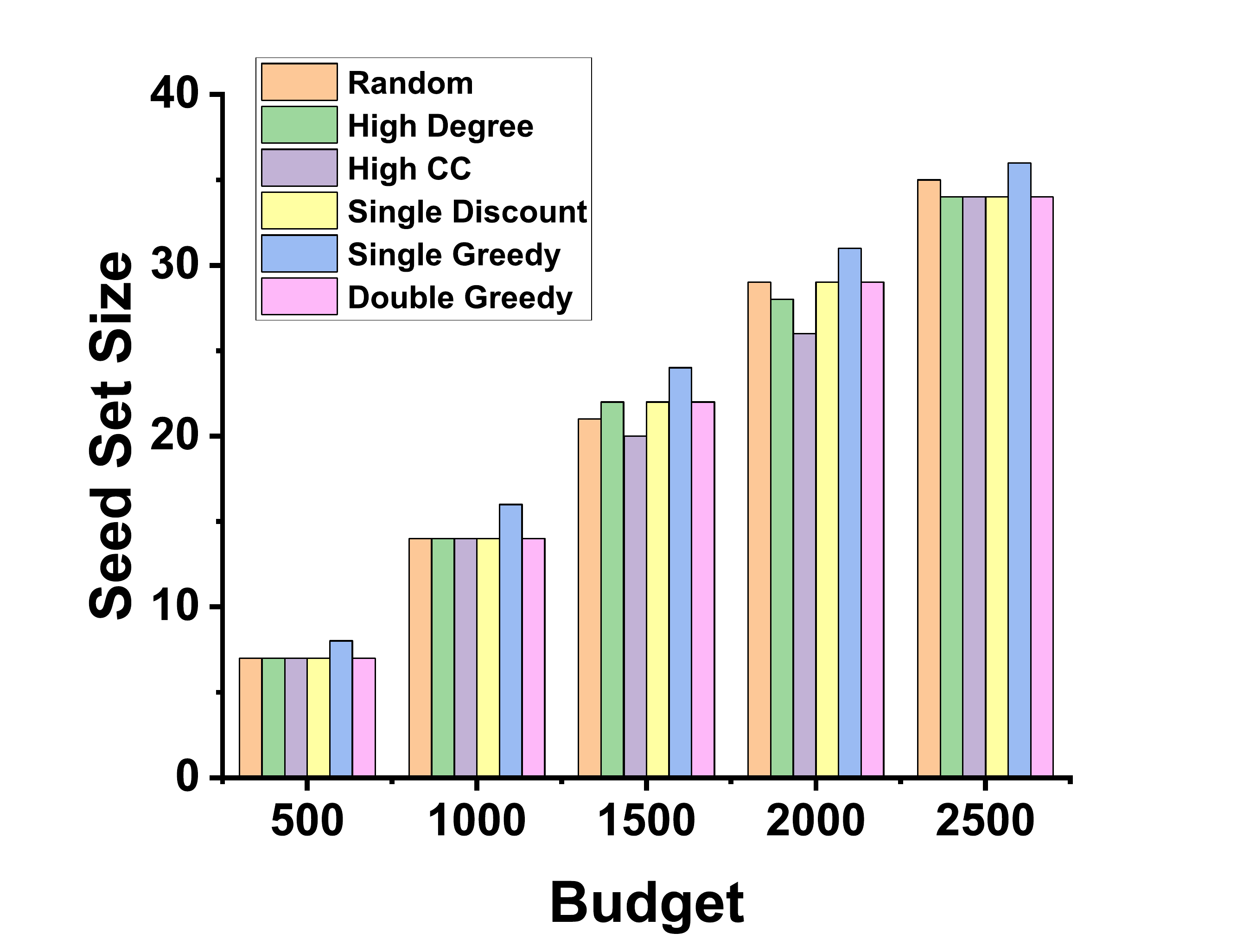} & \includegraphics[scale=0.2]{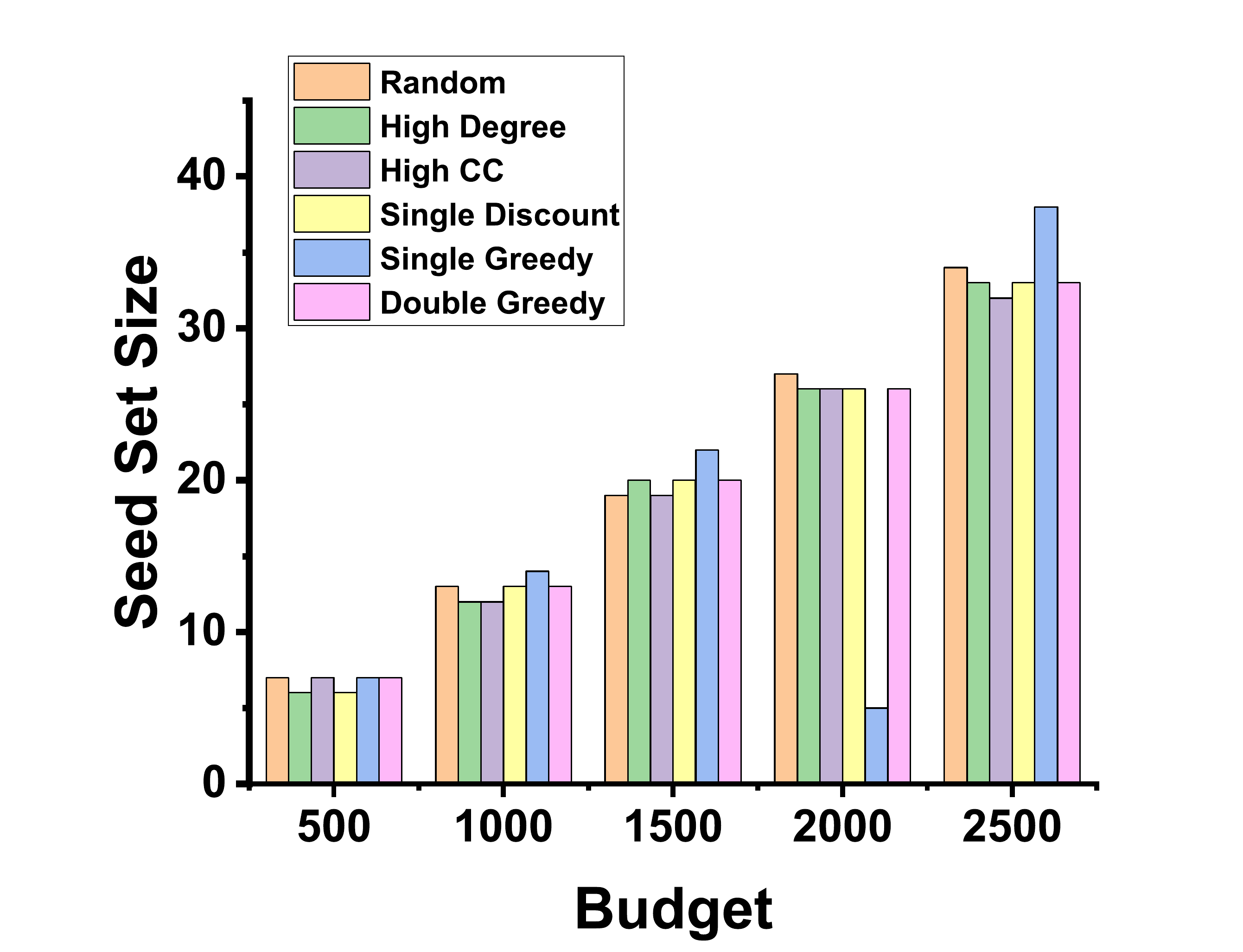} \\
(a) Email-Eu-Core Dataset & (b) Soc-Sign-Bitcoin-Alpha Dataset  & (c) Wiki-Vote Dataset \\
\end{tabular}}
\caption{Budget Vs. Cardinality of the Seed Set Plots in two\mbox{-}phase setting for all the datasets}
\label{Fig:Two_Phase_Seed_Set}
\end{figure*}

\paragraph{\textbf{Algorithms Compared} }
The following algorithms are compared for the performance:
\begin{itemize}
\item \emph {Random}: It randomly selects the seed nodes within the given budget in both the phases. The algortihm is run as per the experimental setup. The algorithm outputs seed set nodes having the largest expected profit.

\item \emph {High Degree}: It selects and evaluates the node with the highest degree. If the node contributes to the non-negative maximum margin, then it is included in the seed set. The algorithm is run as per the experimental setup.

\item \emph {Clustering Coefficient}: The clustering coefficient is the metric to measure the degree to which the nodes in a graph are tend to cluster together. The algorithm selects the node with highest clustering coefficient and evaluates the node for the non-negative maximum margin. The algorithm is run as per the experimental setup.
 
\item \emph {Single Discount}: In single discount the degree of neighbours of the node selected in seed set is reduced by one. The algorithm is as per the experimental setup.

\end{itemize}
\paragraph{\textbf{Results and Discussions} } Now, we describe the experimental results. Figure \ref{Fig:Two_Phase_Seed_Set} shows the budget vs. cardinality of the seed set plots selected by proposed as well as baseline methods over different budget values for different datasets. From this figure, we observe that in most of the problem instances the number of seed nodes selected by the single greedy algorithm is more than the other methods. As an example, for the Email-Eu-Core Dataset when the budget value is $2500$, among the baseline methods the cardinality of the seed sets selected by all of them is $33$. However, in the same setting the number of seed nodes selected by the single greedy approach is $36$. This observation is consistent even for the other two datasets. For the Soc-Sign-Bitcoin-Alpha Dataset when the budget value is $2000$, among the baseline methods the maximum number of seed nodes selected by both Random and Single Discount Heuristics and the number is $29$. However, the same for the single greedy approach is $31$. For the Wiki-Vote Dataset, for the budget value $2500$, among the baseline methods the maximum number of seed nodes selected by Random and the number is $34$ whereas the same by single greedy algorithm is $39$.

\par
\begin{figure*}[!ht]
\centering
\resizebox{\columnwidth}{!}{\begin{tabular}{ccc}
\includegraphics[scale=0.2]{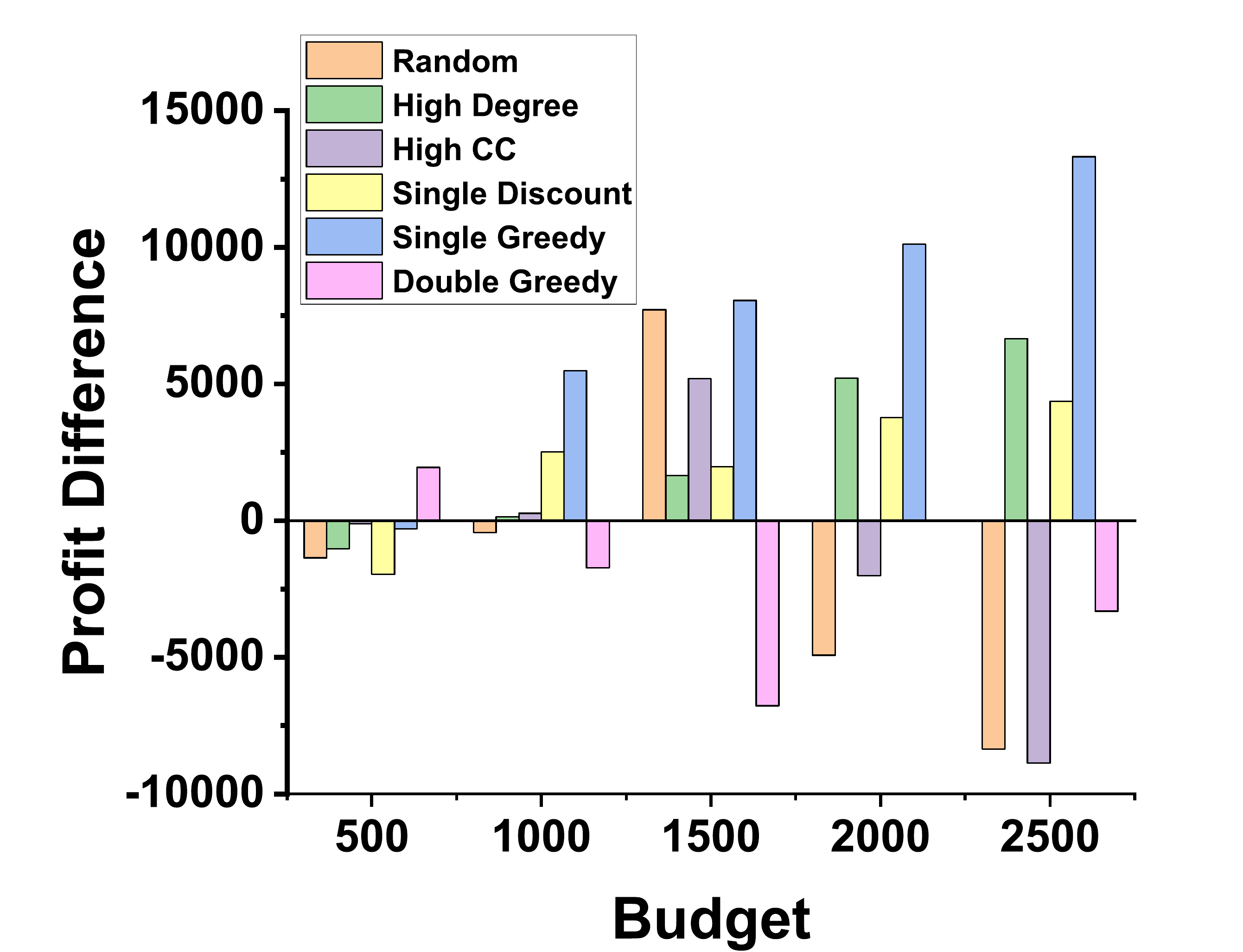} & \includegraphics[scale=0.2]{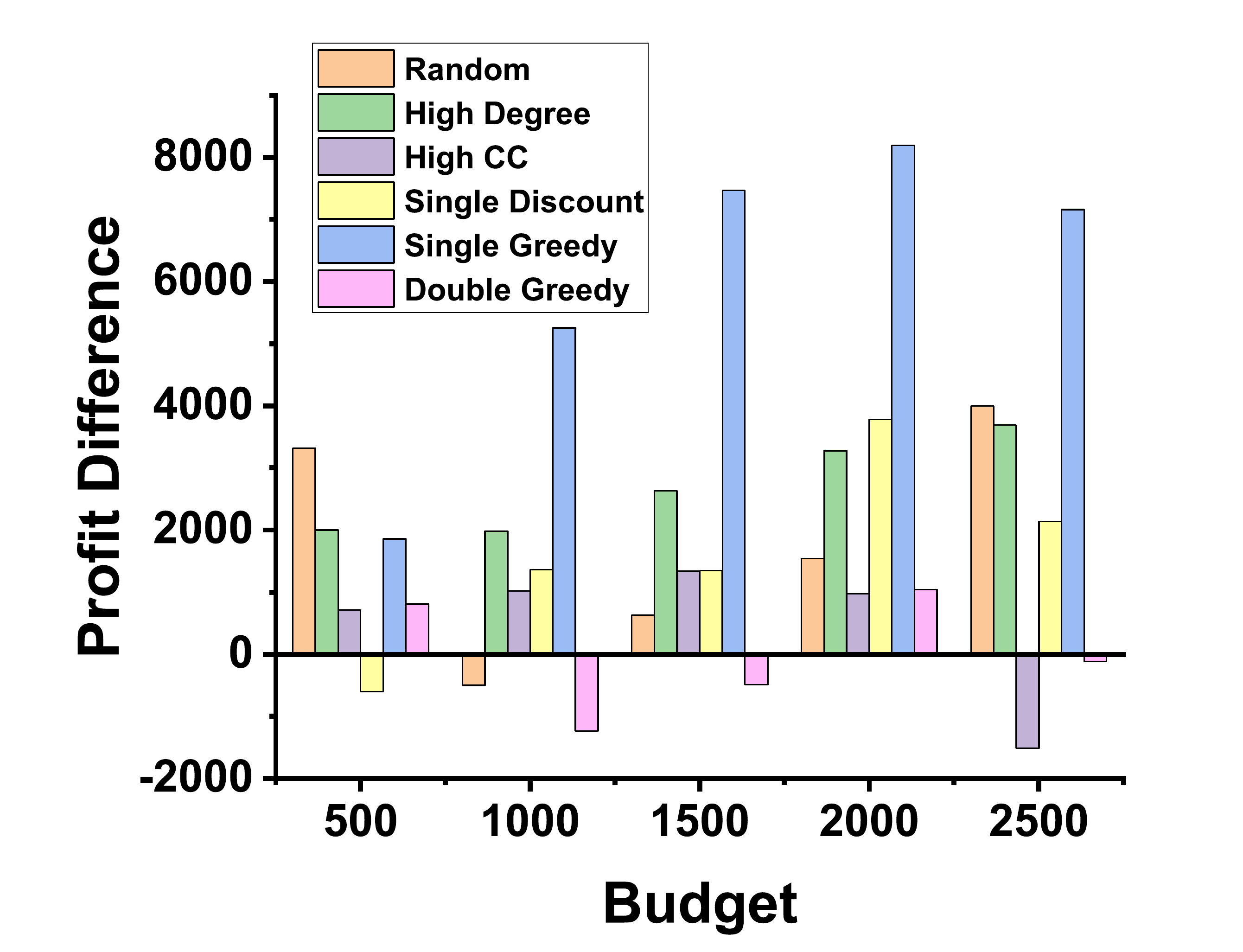} & \includegraphics[scale=0.2]{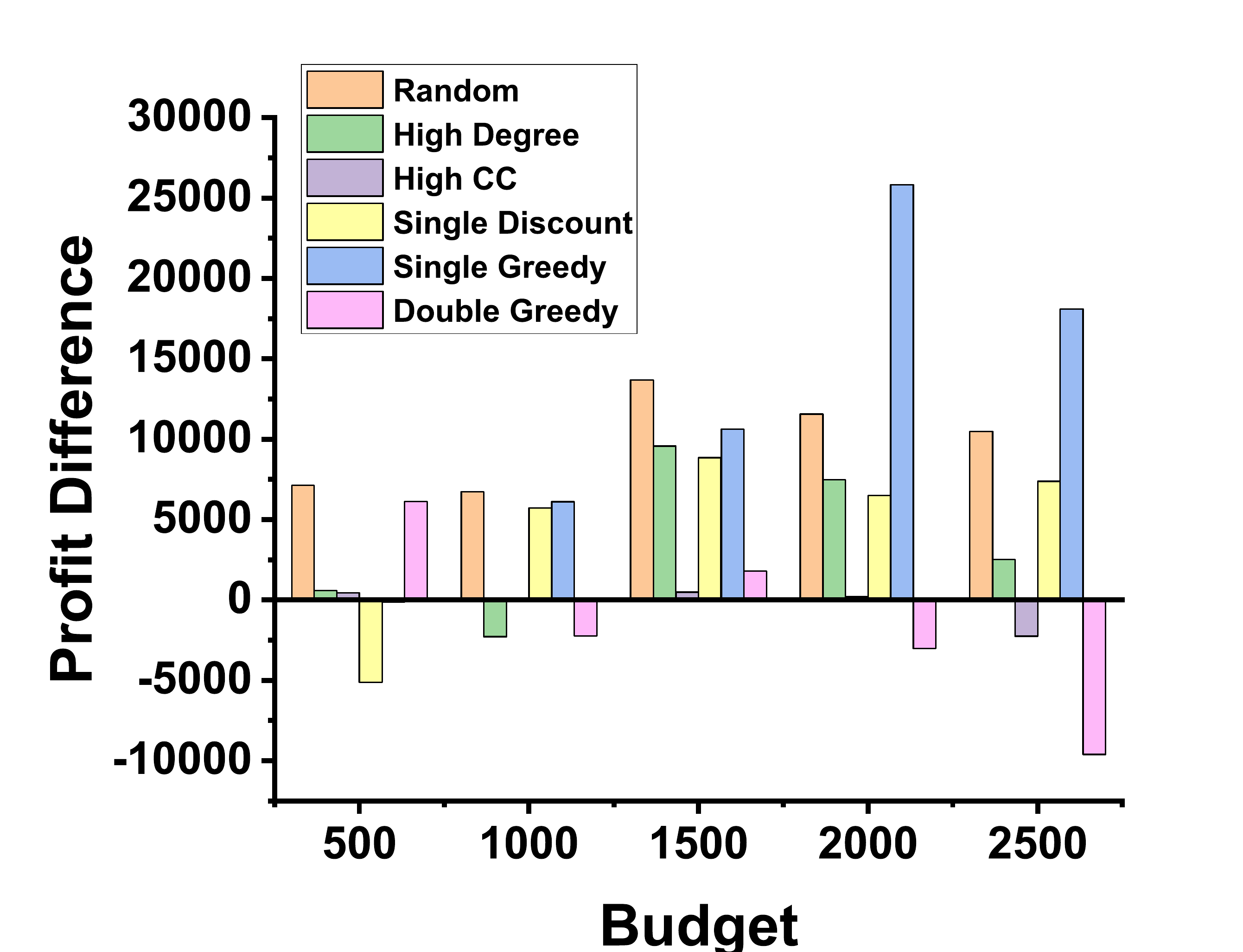} \\
(a) Email-Eu-Core Dataset & (b) Soc-Sign-Bitcoin-Alpha Dataset  & (c) Wiki-Vote Dataset \\
\end{tabular}}
\caption{Budget Vs. Differences in Profit in single and two\mbox{-}phase for all the datasets}
\label{Fig:Profit_Difference}
\end{figure*}

 Figure \ref{Fig:Profit_Difference} shows the budget vs. difference between profit in two\mbox{-}phase and one\mbox{-}phase for different datasets. From the figure, we observe that for most of the budget values the difference in the earned profit remains positive where as for most of the baseline methods this difference is negative. As an example, for the `email-Eu-core' dataset when the budget value is $2500$, for the single greedy algorithm the earned profit in single\mbox{-}phase and two\mbox{-}phase is $56354.35$ and $69676.21$, respectively. Hence, the gain in two\mbox{-}phase is $13158.34$ which is approximately $24 \%$. In general, we observe that when the budget value increases (upto $2000$) this gain increases. However, if we increase the budget further the gain decreases. This is due to the fact when the budget exceeds a threshold, then the earned benefit does not increase much. However, the cost of the seed set increases as the budget increases. Consequently, the profit does not increase even if the budget increases. 
\par Another important observation, we make from Figure \ref{Fig:Profit_Difference} is that the performance of the single greedy and double greedy algorithm is complementary, meaning for the budget values when the single greedy approach does not perform well in those budget values the double greedy approach. As an example, for the Email-Eu-Core dataset for the budget value $500$, the gain in case of the single greedy approach is negative where as the same for the double greedy approach is positive. For the other budget values the scenario is just the reverse.

%

\section{Conclusion and Future Direction } \label{Sec:CFD}
In this paper, we have studied the problem of profit maximization in two\mbox{-}phase setting using online social networks. For this problem, first we develop a mathematical model for this problem based on IC Model of diffusion that captures the profit in an expected sense. Subsequently, we propose two solution methodologies namely single greedy and double greedy. Experimental analysis with real-world datasets show that effectiveness of the proposed solution methodologies. Now, in this study we have not taken care of how the total budget $\mathcal{B}$ is divided into $\mathcal{B}_{1}$ and $\mathcal{B}_{2}$ which we pose for future study.  
%
%
%
 \bibliographystyle{splncs04}
 \bibliography{paper}

\end{document}